\documentclass[a4paper, 11pt]{article}

\usepackage{array}
\usepackage{soul}
\usepackage{natbib}
\usepackage{bm}
\usepackage{multirow}
\usepackage{comment}
\usepackage{longtable}
\usepackage[singlelinecheck=false,labelfont=bf]{caption}
\usepackage{subfloat}
\usepackage{url}

\usepackage{amsmath}
\usepackage{booktabs}
\usepackage{rotating}
\usepackage{graphicx}

\newcolumntype{M}[1]{>{\raggedright\arraybackslash}m{#1}}
\newcommand{\virg}[1]{``#1''}

\providecommand{\keywords}[1]{\vspace*{0.5cm} \noindent \textbf{Keywords:} #1.}

\newlength{\defbaselineskip}
\newcommand{\setlinespacing}[1]%
           {\setlength{\baselineskip}{#1 \defbaselineskip}}

\newcommand{\singlespacing}{\setlength{\baselineskip}{1.5 \defbaselineskip}}

\begin{document}

\title{Validating corruption risk measures: \\a key step to monitoring SDG progress}

\author{Michela Gnaldi$^1$ \and Simone Del Sarto$^1$}

\date{
	$^1$Department of Political Science, University of Perugia -- Via Pascoli 20 06123 Perugia \\ \texttt{\{michela.gnaldi, simone.delsarto\}@unipg.it}
}

\maketitle

\begin{abstract}
The Agenda 2030 recognises corruption as a major obstacle to sustainable development and integrates its reduction among SDG targets, in view of developing peaceful, just and strong institutions.
In this paper, we propose a method to assess the validity of corruption indicators within an Item Response Theory framework, which explicitly accounts for the latent and multidimensional facet of corruption. Towards this main aim, a set of fifteen red flag indicators of corruption risk in public procurement is computed on data included in the Italian National Database of Public Contracts. Results show a multidimensional structure composed of sub-groups of red flag indicators $i.$ measuring distinct corruption risk categories, which differ in nature, type and entity, and are generally non-superimposable; $ii.$ mirroring distinct dynamics 
related to specific SDG principles and targets.

\end{abstract}

\maketitle

\singlespacing
\section{Introduction}\label{sec:intro}
Corruption, the misuse of entrusted power for private gains, is a major obstacle to sustainable development, as it affects each of its five pillars: people, planet, prosperity, peace, and partnerships. Estimates of the costs of corruption show that corruption costs developing and developed countries more than the US\$~10 trillion required to end poverty by 2030~\citep{worldeconomicforum}. These large estimates provide an immediate understanding of the magnitude and effects of corruption in relation to the Sustainable Development Goal (SDG) targets. Resources lost through corruption may instead be used to ensure that everyone has equal access to basic services like health care, education, clean water~\citep{unodc2018}. By diverting resources intended for public goods and services to personal gain, corruption leads to inefficiencies, misallocation of resources, and substandard service delivery. It creates barriers to economic development, by distorting markets and hindering fair competition. It discourages both domestic and foreign investment, limiting economic growth and stifling entrepreneurship. The indirect consequences of corruption are even larger. It undermines the rule of law by distorting legal and regulatory frameworks. It weakens accountability mechanisms and hampers the effective functioning of democratic systems. Corruption dwindles public institutions by compromising their integrity, undermining their effectiveness, and eroding public trust in them.

The adoption of the 2030 Agenda for Sustainable Development was a major breakthrough for the anti-corruption movement~\citep{unodc2018} because it introduced for the first time an explicit link between corruption reduction and peaceful, just and inclusive societies by capitalising on the importance of promoting transparency, accountability and anti-corruption. Indeed, tackling corruption is functional to develop effective, accountable, and transparent institutions, to guarantee efficient allocation of resources, to ensure fair and equal opportunities of access to resources, to boost public institutions and foster trust in them.

Tackling corruption is key to achieving particularly SDG~16, devoted to Peace, Justice and Strong Institutions. Goal 16 has five key targets, which are especially relevant for our research context:

\begin{itemize}
    \item significantly reduce illicit financial and arms flows, strengthen the recovery and return of stolen assets and combat all forms of organised crime 
    (Target~16.4);
    \item substantially reduce corruption and bribery in all their forms 
    (Target~16.5);
    \item develop effective, accountable and transparent institutions (Target~16.6);
    \item ensuring responsive, inclusive, participatory and representative decision making (Target~16.7);
    \item ensure public access to information and protect fundamental freedoms, in accordance with national legislation and international agreements 
    (Target~16.10).
\end{itemize}

The monitoring of these targets is entrusted to a pool of indicators substantially different in nature and capturing, both objectively and subjectively, a specific dimension of a complex and unobservable latent phenomenon~\citep{un2016}. Specifically, the monitoring of Target~16.4 is assigned to two indicators, the total value of inward and outward illicit financial flows (Indicator~16.4.1) and the proportion of seized, found or surrendered arms (Indicator~16.4.2). Direct experiences of corruption and bribery by individuals (Indicator~16.5.1) and businesses (Indicator~16.5.2) in their interactions with public officials are tasked with the monitoring the key Target~16.5. Primary government expenditures as a proportion of original approved budget (Indicator~16.6.1) and the proportion of population satisfied with their last experience of public services (Indicator~16.6.2) are chosen within Target~16.6. Two further indicators are selected for the monitoring of Target~16.7, that is, the proportional representation of various demographic groups (i.e., by sex, age, disability) in national and local institutions, including the legislatures, the public service and the judiciary (Indicator~16.7.1) and the proportion of population who believe decision-making is inclusive and responsive (Indicator~16.7.2). Finally, the monitoring of Target~16.10 is given to the number of verified cases of killing, kidnapping, enforced disappearance, arbitrary detention and torture of journalists (Indicator~16.10.1) and to an indicator measuring the number of countries that adopt and implement a legal framework for public access to information (Indicator~16.10.2).

The choice of these indicators mirrors the  multifaceted
nature of corruption and the complexities of any actions directed at tackling it in view of developing peaceful, just and strong institutions. 
Corruption, like several other phenomena accounted for to improve sustainable development, is a complex latent phenomenon and very challenging to quantify. It encompasses a wide range of economic and social dimensions that are interconnected and can have indirect and long-term impacts not immediately quantifiable.
Indeed, measurement instruments to quantify corruption are wide and diversified and embrace -- other than experience-based measures, such as Indicators~16.5.1 and~16.5.2 included in the SDG~16 -- subjective or perception-based measures, judicial-based measures and statistical inference proxies \citep{gnaldi2021measuring}.
Perception-based measures -- such as the Transparency International's Corruption Perception Index (CPI) -- are widely used in comparative studies,  but are questioned on many grounds. Although CPI scores generally correlate with objective measures such as citizen reported experience with bribery, the literature~\citep{rose2012paying} underlines that perceptions may not reflect factual involvement in corruption crimes. Besides, perceptions can be driven by the general sentiment reflecting prior economic growth~\citep{kurtz2007growth} or media coverage of cases or scandals of corruption~\citep{mancini2017representations}. In recent years, several proposals have been made to identify more \virg{objective} indicators of corruption and overcome the limitations of subjective measures, such as experience-based surveys (such as Transparency International's Global Corruption Barometer), experimental approaches, conviction rates~\citep{fiorino2012corruption}, context analyses (e.g. newspapers and press reports), and proxy approaches~\citep[e.g.,][]{golden2005proposal, olken2007monitoring}, including corruption risk indicators.

In this paper, we focus on these latest generation of corruption indicators, the so-called red flag indicators of corruption risks, and propose a method to assess their degree of validity. Red flags are proxy measures for corruption signalling risks of corruption, rather than actual corruption. They are expected to be correlated with corrupt practices, rather than perfectly matching them \citep{fazekas2020uncovering, oecd2019}.  

Red flags, as well as many indicators employed to measure sustainable development, can be interrelated. The presence of one red flag may indicate a higher likelihood of other associated risks, just as progress in one area can have implications for others when measuring sustainable development progress. Capturing these interlinkages and understanding the complex relationships among red flags require careful analysis and measurement frameworks that account for systemic relationships among them. Besides, red flags -- as well as many other indicators employed to measure SDG targets -- often involve subjective judgements and interpretation. Determining which factors or behaviours constitute red flags and how they should be assessed requires expert judgement and contextual understanding.

In line with the previous considerations, in this paper we propose a method to assess the degree of validity of red flags for corruption risk with respect to their internal coherence and based on a procedure grounded on the framework of multidimensional Item Response Theory (IRT) models (specifically, by means of the multidimensional graded response model).
This framework is particularly appropriate for the context at issue as it allows us to study the extent to which corruption measures capture the higher-level theoretical construct by assuming that the associations among items (i.e., red flags) are represented by a latent and unobservable trait (corruption).

The article is organised as follows. In Section~\ref{sec:redflags} we go into greater detail about red flag indicators and discuss the ways their validity is addressed in the literature and in the present paper. Section~\ref{sec:irt} is devoted to introduce the IRT framework and to describe the specific IRT model used in our case study, whereas Section~\ref{sec:validity_results}, after introducing the data at hand and the selected red flag indicators, reports the results of the dimensionality assessment, whose discussion is provided in Section~\ref{sec:concl} together with the main conclusions.

\section{Red flag indicators of corruption risk}\label{sec:redflags}
In this section we further explore the broad framework of red flag indicators of corruption risk (Section~\ref{subsec:redflags_general}) and examine how the literature and this paper tackle the issue of their validity (Section~\ref{subsec:redflags_validity}).
\subsection{General framework}\label{subsec:redflags_general}
International experts and policy makers who work in the field of corruption have come to recognise over the past 20 years that prevention and repression are both important components of effective corruption control~\citep{carloni2017misurare, gnaldi2019corruption, gnaldi2021measuring, gallego2021preventing}. When a corrupt act is committed, repression steps in to punish it. Differently, corruption prevention seeks to identify and reduce opportunities for corruption. Concurrently, the so-called red flags of corruption risk, a new generation of indicators that are thought to be more accurate and effective than pre-existing measures, have been developed~\citep{mungiu2016new}. Instead of \textit{ex-post} assessing the \virg{amount of corruption}, the primary goal of such metrics is an \textit{ex-ante} identification of situations potentially vulnerable to corruption.

In order to identify corruption risks and establish practical mitigation solutions, corruption risk assessment systems are progressively built around corruption red flags. They have been increasingly empowered by both data availability in a machine-readable format and the development of new technologies based on the collection and cross-processing of public data sources \citep{gnaldi2021measuring}. For example, the reader can refer to the system set up by the Italian anticorruption authority (ANAC) within the project \virg{Measurement of corruption risk at territorial-level and promotion of transparency}\footnote{\url{https://www.anticorruzione.it/il-progetto}}.

Red flags are typically used to measure the likelihood of corruption in the public procurement process, through which public authorities acquire goods, products, or services from businesses. A number of occurrences in the public procurement process -- such as a delay or failure to award a contract, an overuse of emergency procedures, contract extensions, the repetition of small assignments for the same item, etc. -- can indicate potentially risky scenarios to watch out for. In fact, one of the areas of government where corruption is most likely to occur is public procurement. Every stage of the procurement process is susceptible to integrity concerns \citep{oecd2016}, which are aggravated by the economic volume of transactions and involved financial stakes. 

Corruption in public procurement aims at directing the contract secretly and repeatedly toward the preferred bidder. The primary aim of institutionalised grand corruption is rent extraction. Rents can be obtained in public procurement by pre-selecting businesses, who then make additional profit by charging higher than average market prices for the contract object delivery \citep{abdou2021covid}. Information on the price and amount of procured deliveries -- which is typically present in most administrative public procurement databases, though not comparable over time and space -- would be needed in order to assess corruption risk by quantifying extra-profit.

As a result, other indicators of corruption risk are put forth, such as the circumstance that a procurement procedure receives a single offer by a single bidder. Indeed, it has been demonstrated that the number of bids submitted, particularly when there is only one submitted bid, is related to the likelihood of corruption \citep{klavsnja2015corruption}, and this relationship has been widely used in the literature as a proxy for corruption. Additional frequent risk indicators are \citep{fazekas2020uncovering}: non-open or exceptional procedure forms, such as direct awards, that allow for the suppression of competition and the steering of contracts to the chosen bidders \citep{auriol2016public}; little time given for bid advertisements, which may prevent unaffiliated bidders from developing suitable offers in time; absence of publication of the call for tenders (where publishing is voluntary); short time intervals to award contracts, as hasty decisions may indicate preconceived evaluation; subjective and difficult-to-quantify evaluation criteria (used in place of price-related evaluation criteria), which indicate some discretionary margins and may restrict accountability management systems.
\subsection{The validity issue}\label{subsec:redflags_validity}
The validity issue of red flag indicators of corruption risk is as strategic to monitor SDG achievements as it is complex and still largely unexplored in the international scientific literature. Indeed, the validity of a measurement tool concerns, in general terms, its ability to adequately reflect the concept to be measured \citep{adcock2001measurement}. Thus, corruption measures are  valid when they detect corruption precisely in the cases where there has actually been corruption. But the evidence that there really was corruption is not straightforwardly inferable as corruption is a hidden and latent phenomenon. Recourse to final judgements for bribery offences would intuitively be the most direct way to assess the degree of validity of corruption measures. However, resorting to sentences presents various substantial and technical problems attributable to the fact that they detect only a very marginal part of emerged corruption. Moreover, judgements are expressed qualitatively and are included in documents whose format may not be easily accessible and directly analysable but with data scarping techniques and text mining methods. The latest would be needed to access basic information, for instance, on the type of crime, the actors of the offence (i.e., individuals, companies, public officials, contracting authorities, etc.), the period in which the offence was committed, etc. Furthermore, judgements are usually issued many years after the crime has been committed. Given this temporal misalignment, a retrospective evaluation of the validity of corruption measures based on them presupposes an efficient judicial system capable of keeping track of the crime actors, type of crime, etc. over time, other than a system which allows interested users to access freely judgement contents. 

Validating corruption risk indicators is rarely addressed by the scientific literature on corruption measurement, with only a few exceptions. In \cite{fazekas2018innovations}, red flags are validated with respect to their strength in predicting the single bidding indicator within a logistic regression modelling framework. In this procedure, the single offer -- which measures the circumstance that a public contract receives only one bid -- is the key criterion against which to verify the validity of a set of red flags. The procedure has the merit of being easily replicable, as the single offer indicator can be straightforwardly computed with data accessible in most Countries. On the other hand, it implicitly assumes that when a public contract receives only one bid we are faced with a certainly corrupt circumstance, an occurrence which, though, would need to be ascertained as well. 

Another well-known procedure is the one described in \cite{decarolis2019corruption} and \cite{decarolis2022corruption}. Employing maximum likelihood algorithms (i.e., LASSO, ridge regression and random forest), the authors suggest evaluating the reliability of red flags in terms of their ability to accurately predict a variable indicating whether a company's owners or top executives have ever been under investigation by the police for engaging in corruption practices. Despite the strength of the variable expressing corruption risk and derived from police investigations, the procedure appears limited in its replicability, as police investigation data are confidential and hardly open.

Given the complexity and limitations of validating corruption measures relying on the above procedures, in this paper we opt for the adoption of a criterion of \virg{minimum validity} \citep{bello2021measuring}, which leads to assessing the degree of validity of a set of indicators with respect to their level of internal coherence, based on specific statistical criteria.
According to such a criterion, corruption measures  can be valid when they are strongly correlated to the latent phenomenon they measure. Specifically, the interest is to assess to what extent corruption measures can capture the higher-level theoretical construct they intend to detect, while at the same time excluding irrelevant elements. 

Coherently with the criterion above, in this paper we assess the validity of a set of red flag indicators though a procedure based on the framework of multidimensional Item Response Theory (IRT) models. By assuming that the associations between items (i.e., red flags) are represented by a latent and unobservable trait (corruption), this framework is especially suitable for the research context at hand because it enables us to investigate the degree to which corruption measures capture the higher-level theoretical construct. Besides, in their multidimensional version, IRT models assume that the multidimensional latent trait is measured by a multivariate latent variable. This latest extension is, again, very suitable for the purposes of this study as it accounts for the complexity of corruption, that is, for the circumstance that the relational structure among single indicators might mirror a complex and multidimensional configuration, rather than a simple and unidimensional one, where high correlations can be observed between subgroups (or sub-dimension) of red flags.  
\section{Item Response Theory models to evaluate the validity of red flags}\label{sec:irt}
In this section we first introduce the IRT framework from a general point of view (Section~\ref{subsec:irt_prel}), then the specific model used for the analysis (i.e., the graded response model) is described (Section~\ref{subsec:model}).
\subsection{Preliminaries}\label{subsec:irt_prel}
IRT models are latent variable models for handling multivariate categorical data, assuming that associations among items (i.e., red flags in our case) may be represented by a latent trait (corruption risk here), which can be uni- or multidimensional, that is, measured by a uni- or multivariate latent variable~\citep{bartolucci2015statistical}. IRT is broadly used in analysing students' proficiency through learning assessment test, where the items (test questions) are binary variables (wrong/correct responses). In this regard, Rasch~\citep{rasch} and two-parameter logistic~\citep[2PL;][]{2pl} models are the most widely known and employed IRT models. However, in other research fields items can have more than two categories, therefore IRT for polytomous items are available (based on nominal or ordinal items). 

Similarly to factor analysis for continuous variables, IRT can be employed for confirmatory and exploratory analysis of the dimensionality structure of the underlying latent variables when the manifest variables are categorical.
Data dimensionality investigation is an essential step towards the validation of corruption risk indicators. Indeed, as clarified throughout this paper, the measurement of the risk of corruption through proxy indicators is valid to the extent that red flag indicators are strongly related to the latent concept (i.e., corruption) they intend to measure. When the phenomenon of interest is a one-dimensional latent variable, it is expected that the proxy indicators used for its measurement, in order to be valid, are all strongly correlated with each other, as an expression of a single underlying one-dimensional latent. Instead, when the latent phenomenon is complex and multidimensional -- such as the risk of corruption -- it is reasonable to expect that the red flag indicators are not all related with each other, but rather in sub-groups of indicators, each measuring a different sub-dimension and aspect of the same underlying risk. 

IRT models are particularly appropriate methodological tools in the context at issue as they allow us to study the relational structure of single red flags (i.e., their dimensionality) and can be applied when one has no prior information about the number and composition of the sub-dimensions of the phenomenon under scrutiny. In the IRT case, the dimensionality structure can be estimated empirically by comparing nested models and rotating ascertained solutions to seek for more interpretable structures \citep{bock1981marginal, bock1988full, chalmers2012mirt}. 

Moreover, the choice of this approach for evaluating the dimensionality structure of the red flags has been driven by the nature of the available data. As outlined in Section~\ref{subsec:exploratory_analysis}, our set of red flags, computed at the contracting authority level, displays skew distributions of varying intensities and cannot be considered  normally distributed. As a consequence, any methods based on the normal distribution assumption is inappropriate for the data at hand and we opt for an IRT modelling framework. Specifically, we consider an IRT model for handling ordinal items, the graded response model (GRM) of \cite{samejima1969estimation} -- introduced in the next section -- and we convert our elementary indicators into categorical (ordinal) indicators using a unified scale based on four categories, representing distinct levels of corruption risk.

\subsection{The multidimensional graded response model}\label{subsec:model}
The multidimensional GRM proposed by~\cite{samejima1969estimation} can be formalised as follows. Let $Y_{ij}$ be the value of indicator $j$ as regards contracting authority $i$, with $i=1,\ldots,n$ and $j=1,\ldots,J$. Let us assume that each indicator is ordinal with $C_j$ categories (labelled as 0, 1, \ldots, $C_j-1$). In our setting, each indicator has the same number of categories ($C=4$). For modelling ordinal responses and assuming multiple underlying latent traits, the multidimensional version of the GRM can be used and consists in the natural extension of the 2PL multidimensional IRT model for polytomous data.

Let us denote the multidimensional latent trait of contracting authority $i$ by $\bm{\theta}_i=[\theta_{i1}, \ldots, \theta_{iD}]^\top$, which represents its unobservable corruption risk in public procurement with $D$ dimensions (or factors). Like for its unidimensional counterpart, the multidimensional GRM formulation starts with the definition of the (conditional) probability that indicator $j$ of contracting authority $i$ is observed in category $y$, $P(Y_{ij} = y \vert \bm{\theta}_i)$, which can be written as the difference of two cumulative probabilities:
\[
P(Y_{ij} = y \vert \bm{\theta}_i) = P^*_{ij}(y) - P^*_{ij}(y+1),
\]
\noindent where $P^*_{ij}(y)$ is the probability of observing category $y$ or higher in indicator $Y_{ij}$, namely $P^*_{ij}(y)=P(Y_{ij} \geq y \vert \bm{\theta}_i)$, with $y=0,\ldots,C-1$.

Cumulative probability $P^*_{ij}(y)$ can be written using two parameters related to the indicator, as follows:
\[
P^*_{ij}(y) = \frac{\exp(\bm{a}_j^\top \bm{\theta}_i + b_{jy})}{1+\exp(\bm{a}_j^\top \bm{\theta}_i + b_{jy})},\qquad y=1,\ldots,C-1, 
\]
\noindent where $\bm{a}_j=[a_{j1}, \ldots, a_{jD}]^\top$ is the vector of $D$ slopes of indicator $j$ (one for each dimension) and $b_{jy}$ is the intercept of indicator $j$ with reference to category $y$. Obviously, cumulative probability for the first response category ($y=0)$ is not formally defined, as $P^*_{ij}(0)=1$. 

As we can see, for each indicator the model envisages as many slopes as the number of dimensions $D$, so as to give the indicator a chance, in theory, to \virg{load} on all the dimensions. However, in order to ease the result interpretation, slopes can be transformed in a sort of loadings (like in factor analysis) and successively suitably rotated in order to reach a dimensional structure as simple as possible. Specifically, the loading of item $j$ on dimension $d$ is obtained as
\[
l_{jd} = \frac{a_{jd}}{\sqrt{1+\sum_d a_{jd}^2}},
\]
while other typical measures of factor analysis can be computed, such as communality $h^2_j$, uniqueness $u_j$ and SS loadings $SS_d$, as follows:
\begin{align*}
    h^2_j &= \sum_d l^2_{jd},\\
    u_j &= 1 - h^2_j, \\
    SS_d &= \sum_j l^2_{jd}.
\end{align*}
Moreover, indicator parameters can also be expressed with reference to difficulty and discrimination, as typically done in the IRT framework~\citep{reckase2009}. In particular, components of $\bm{a}_j$ can be considered as measures of the indicator discrimination with respect to each dimension. However, a measure of overall discrimination of item $j$ over the $D$ dimensions (denoted by $\alpha_j$) can be computed as follows:
\begin{equation}\label{eq:overall_discr}
    \alpha_j = \sqrt{\sum_d a^2_{jd}}.
\end{equation}
Similarly, intercepts can be converted into overall difficulties as follows:
\begin{equation}\label{eq:overall_diff}
    \beta_{jy} = \frac{-b_{jy}}{\alpha_j}, \qquad y=1,\ldots,C-1.
\end{equation}
In particular, $\beta_{jy}$ is the overall level of latent trait at which the probability of observing category $y$ or higher in indicator $j$ is equal to that related to category $y-1$ or smaller.

Model estimation procedure using maximum likelihood methods is detailed in~\cite{bock1988full} and implemented in the R package \virg{mirt}~\citep{chalmers2012mirt}. 
\section{Validity assessment of red flag indicators through the multidimensional graded response model}\label{sec:validity_results}
This section is devoted to detail the main results of the paper about the validation of red flag indicators. After describing the data at hand and the selected red flag indicators in Section~\ref{subsec:data_indicators}, an exploratory analysis of these data is presented in Section~\ref{subsec:exploratory_analysis}, whereas the final results on the validity of the red flag indicators using the graded response model is given in Section~\ref{subsec:irt_solutions}.
\subsection{Data and indicators}\label{subsec:data_indicators}
For this work, we rely on a pool of red flag indicators computed at the contracting body level, by aggregating information of the contracts managed by each contracting authority in the considered time period. Every tender managed by Italian contracting authorities flows into the Italian National Database of Public Contracts (BDNCP for short, standing for \textit{Banca Dati Nazionale dei Contratti Pubblici}). It is managed by the Italian anticorruption authority and its open version is organised in several tables, one for each stage of the procurement process \citep{anac2023}. In the ANAC open data catalogue\footnote{\url{https://dati.anticorruzione.it/opendata}}, users can download any specific table about a particular procurement stage (in csv or json format) and including specific information on the contracts that have reached the selected phase. For example, the table on award notices (\textit{Aggiudicazioni}) includes information on  awarded contracts, such as award notice date, award value, award criterion, number of received and eligible bids, etc.

Given the above, it is not possible to get the entire BDNCP by a one-click download as users need to download the single tables and then merge them. Two keys are available to this aim: $i.$ the contract unique identifier, called CIG (\textit{Codice Identificativo Gara}), assigned to each procedure during the call for tenders phase; $ii.$ the award ID, obtained once the contract reaches the award phase and allowing us to retrieve information on the award notice subsequent stages.

Tables~\ref{tab:sel_redflags1}, \ref{tab:sel_redflags2} and \ref{tab:sel_redflags3} list the fifteen selected red flags and report their label, the procurement stage, some formalisation details and the variables needed for their computation. Moreover, the reader can refer to Appendix~\ref{appendix} for a detailed description of each red flag.

\begin{subtables}\label{tab:redflags}
    \begin{sidewaystable}
        \caption{Selected red flags (1-6)}
        \label{tab:sel_redflags1}
        \footnotesize
        \begin{tabular}{p{1.8cm}*{6}{M{0.12\textwidth}}}
            \toprule
            & \textbf{Red flag 1} & \textbf{Red flag 2} & \textbf{Red flag 3} & \textbf{Red flag 4} & \textbf{Red flag 5} & \textbf{Red flag 6} \\ 
            \\
            & Proportion of non-open procedures (number) & Proportion of non-open procedures (value) & Proportion of procedures with a single bid (number) & Proportion of procedures with a single bid (value) & Proportion of procedures awarded using MEAT criterion (number) & Proportion of procedures awarded using MEAT criterion (value) \\
            \midrule
            \textbf{Label} & \textit{non\_open\_count} & \textit{non\_open\_val} & \textit{single\_bid\_count} & \textit{single\_bid\_val} & \textit{MEAT\_count} & \textit{MEAT\_val}  \\
            \midrule
            \textbf{Procurement stage} & Publication of the call for tenders & Publication of the call for tenders  & Award notice & Award notice & Award notice & Award notice \\
            \midrule
            \textbf{Formalisation details} 
            & \textit{Numerator}: number of \virg{non open} contracts; \newline \textit{Denominator}: total number of contracts 
            & \textit{Numerator}: total value of \virg{non open} contracts; \newline \textit{Denominator}: total value of all the contracts 
            & \textit{Numerator}: number of contracts receiving a single bid;\newline
           \textit{Denominator}: total number of awarded contracts 
           & \textit{Numerator}: total value of contracts receiving a single bid; \newline \textit{Denominator}: total value of awarded contracts 
           & \textit{Numerator}: number of contracts awarded with MEAT; \newline \textit{Denominator}: total number of awarded contracts 
           & \textit{Numerator}: value of contracts awarded with MEAT; \newline \textit{Denominator}: total value of awarded contracts \\
            \midrule
            \textbf{Required variables} & Contract typology (open or non open) & Contract typology (open or non open) and value & Number of received bids, number of eligible bids & Number of received bids, number of eligible bids, contract value & Award criterion & Award criterion, contract value \\
            \bottomrule
        \end{tabular}
    \end{sidewaystable}
    
    \begin{sidewaystable}
        \caption{Selected red flags (7-11)}
        \label{tab:sel_redflags2}
        \footnotesize
        \begin{tabular}{p{1.8cm}*{5}{M{0.15\textwidth}}}
        \toprule
        & \textbf{Red flag 7} & \textbf{Red flag 8} & \textbf{Red flag 9} & \textbf{Red flag 10} & \textbf{Red flag 11} \\ 
        \\
        & Average advertisement period 
        & Average bid evaluation period 
        & Proportion of excluded bids 
        & Proportion of procedures with all bids excluded but one
        & Proportion of excluded bids in procedures with all bids excluded but one \\
        \midrule
        \textbf{Label} & \textit{advertisement} & \textit{evaluation} & \textit{excluded\_bids} & \textit{all\_bids\_excluded\_but\_one} & \textit{excluded\_bids\_but\_one}  \\
        \midrule
        \textbf{Procurement stage} & Publication of the call for tenders & Award notice  & Award notice & Award notice & Award notice  \\
        \midrule
        \textbf{Formalisation details} 
        & Average time distance between publication of the call for tenders and bid submission deadline  
        & Average time distance between bid submission deadline and award notice
        & Average proportion of excluded bids out of the total number of bids received (over the awarded contracts) 
        & \textit{Numerator}: number of contracts in which all bids but one are excluded; \newline \textit{Denominator}: total value of awarded contracts 
        & Average proportion of excluded bids out of the total number of bids received (over the contracts with all bids excluded but one) \\
        \midrule
        \textbf{Required variables} & Date of bid submission deadline and publication of the call for tenders 
        & Date of award notice and bid submission deadline 
        & Number of excluded bids  
        & Number of received bids, number of excluded bids 
        & Number of received bids, number of excluded bids \\
        \bottomrule
        \end{tabular}   
    \end{sidewaystable}

    \begin{sidewaystable}
        \caption{Selected red flags (12-15)}
        \label{tab:sel_redflags3}
        \footnotesize
        \begin{tabular}{p{1.8cm}*{4}{M{0.19\textwidth}}}
        \toprule
        & \textbf{Red flag 12} & \textbf{Red flag 13} & \textbf{Red flag 14} & \textbf{Red flag 15} \\ 
        \\
        & Proportion of procedures with at least one modification 
        & Average deviation in contract cost
        & Average deviation in contract duration
        & Winners' distribution homogeneity \\
        \midrule
        \textbf{Label} & \textit{modifications} & \textit{amount\_deviation} & \textit{time\_deviation} & \textit{winners\_homog}  \\
        \midrule
        \textbf{Procurement stage} & Ending  & Ending  & Ending & Award notice  \\
        \midrule
        \textbf{Formalisation details} 
        & \textit{Numerator}: number of contracts with at least one variant; \newline \textit{Denominator}: total number of concluded contracts    
        & Average ratio between contract award value and final sums paid 
        & Average ratio between real and expected contract duration  
        & Degree of homogeneity of the frequency distribution of winner firms for a particular contracting body  \\
        \midrule
        \textbf{Required variables} & Contract ending, occurrence of variants 
        & Award value, paid off value  
        & Expected and effective contract ending date   
        & Winner ID  \\
        \bottomrule
        \end{tabular}        
    \end{sidewaystable}
\end{subtables}
\subsection{Exploratory analysis}\label{subsec:exploratory_analysis}
In this work, we consider data from the BDNCP about tenders with a call published in 2017, for a total number of more than 145 thousand contracts. The choice of the referenced year is justified by the need to include in the analyses an adequate portion of recent contracts reaching the final stages of the procurement process, whose average duration is five years. 

According to the structure of the BDNCP open version (as outlined in Section~\ref{subsec:data_indicators}), the final data matrix, which pertains to the portion of selected contracts, has manually been created as follows. Firstly, the list of contracts published in the selected year is obtained by downloading the related table about the call for tenders (\textit{Bando CIG} in the ANAC open data catalogue). Then, using the contract unique identifier (the CIG), it is merged with the award notice table (\textit{Aggiudicazioni}), hence connecting each contract with its own award ID. This secondary key is then used for retrieving the information of each contract (if available) related to the other procurement stages, required for the computation of the selected red flag indicators. Specifically, the tables about the following stages are considered: variants (\textit{Varianti}), contract start (\textit{Avvio contratto}), contract end (\textit{Fine contratto}), contract economic framework (\textit{Quadro economico}), and winners (\textit{Aggiudicatari}).

Due to contract specific features and/or missing values, a sub-sample of 7,900 contracting bodies has been accounted for in our study. Missing values can occur due to several reasons. A major missing case occurs when contracting bodies manage a single procedure over a year, a quite typical circumstance for \virg{small} bodies. Anytime this condition applies, the computation of the red flag at the contracting body level is omitted, as (at least) two contracts are needed. Missing values in an indicator may also occur when the information required for the computation of an indicator is not present in the database. In this case, the indicator cannot be calculated due to missing variable(s), even if the procurement procedure has got the formal characteristics that would allow us to compute it.

\begin{figure}
    \centering
    \includegraphics[trim={6cm 0 9cm 0}, clip, width=\textwidth]{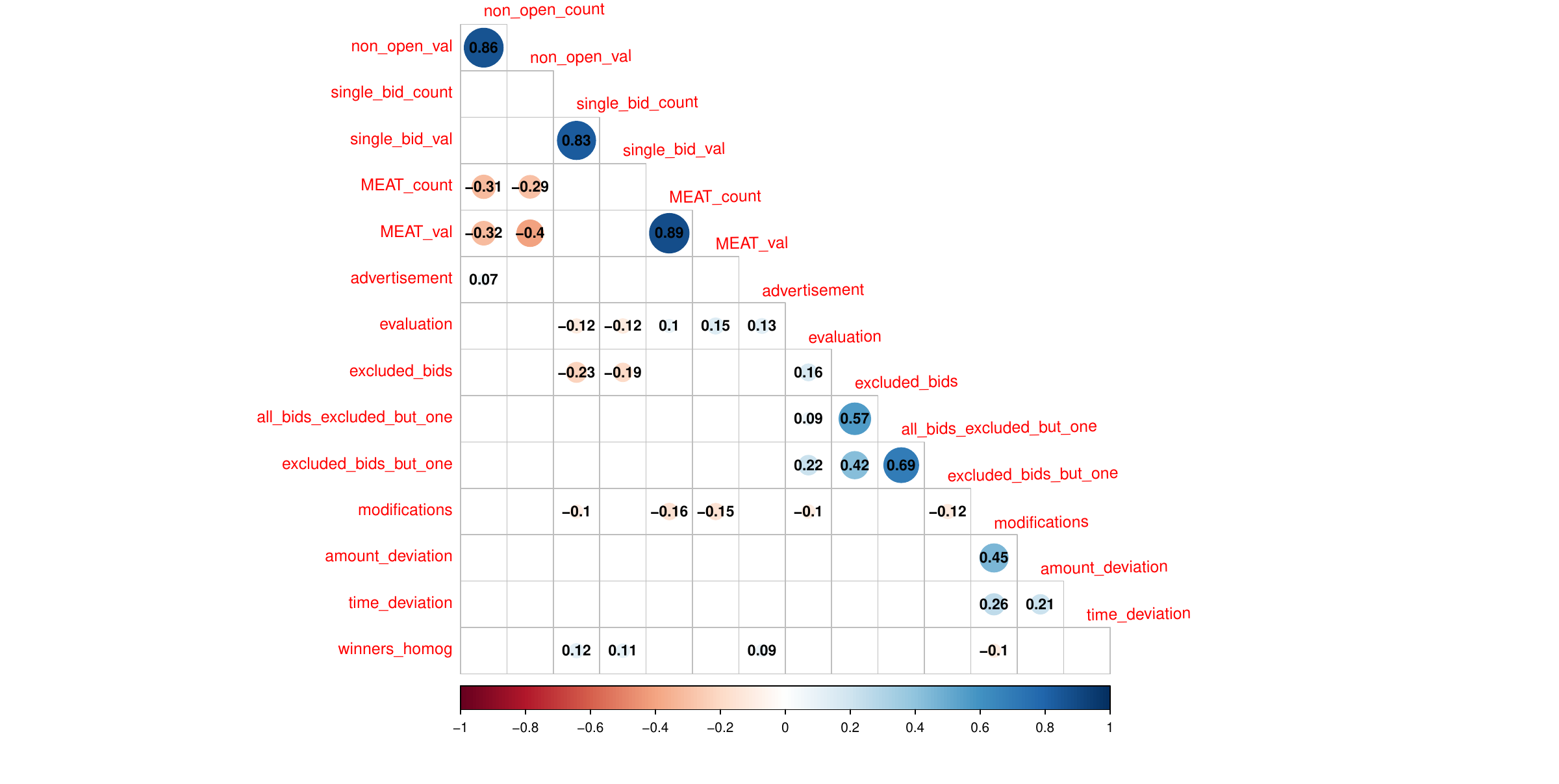}
    \caption{Linear correlation plot among red flags. Note: only significant correlations are reported, adjusted for multiple comparisons.}
    \label{fig:corrplot}
\end{figure}

As a first exploratory tool to investigate the dimensionality of the selected red flags, we run an analysis of the correlation between pairs of indicators, relying on the Pearson's linear correlation and the Spearman's rank correlation coefficients. In this regard, we can observe from Figure~\ref{fig:corrplot} the Pearson correlation plot between the fifteen indicators. Only significant correlations are shown, where the significance level is set at 5\% and adjusted for multiple comparisons (105, in this case). Very similar results are obtained by means of the rank correlation and for this reason we omit to report them.

As expected, the red flags belonging to the first six indicators are highly and positively correlated within pairs since each pair accounts for two facets of a mutual risk. In fact, the correlation coefficient is 0.86, 0.83 and 0.89 for indicator related to non open procedures (having root label \textit{non\_open}), single bidding (\textit{single\_bid}) and usage of MEAT criterion (\textit{MEAT}), respectively. In addition, the two indicators on the use of non-open procedures (\textit{non\_open\_count} and \textit{non\_open\_val}) are negatively correlated with indicators about the use of MEAT as award criterion (\textit{MEAT\_count} and \textit{MEAT\_val)}, albeit with a small extent (between 0.3 and 0.4 in absolute value). Moreover, two other groups of three connected red flags emerge from the correlation plot: a first group includes the indicators accounting for the exclusion of bids (\textit{excluded\_bids}, \textit{all\_bids\_excluded\_but\_one} and \textit{excluded\_bids\_but\_one}), while the second, less evident (i.e., with lower correlations), contains red flags concerning procedures with variants (\textit{modifications}), deviations in the contract economic value  (\textit{amount\_deviation}) and duration (\textit{time\_deviation}).

In our data, the red flags at the contracting authority level, expressed as continuous variables, show distributions with different degrees of skewness (data not shown). Hence, they cannot be treated as normally distributed variables and a factor analysis approach cannot be considered a proper solution for exploring their dimensionality \citep{gao2008nonnormality, mahmoud2015confirmatory, mahmoud2019validating, meyers2016applied}.

Furthermore, given that the red flags are expressed either as proportions or as means, we transform them into ordinal indicators expressed into a common scale with the same categories, each representing a different degree of corruption risk.
Specifically, the fifteen selected red flags are discretised into four categories of increasing corruption risk, labelled from 0 to 3, where 0 stands for \virg{no or low risk}, 1 for \virg{medium-low risk}, 2 for \virg{medium-high risk} and 3 for \virg{high risk}. 
Since there is no general and objective recommendation on how to identify corruption risk categories, thresholds are computed by relying on the quantiles of each indicator observed distribution. 

With the only exception of the red flags concerned with the time for submitting bids and for selecting the winning bids (\textit{advertisement} and \textit{evaluation}), all other indicators have a positive polarity, so that the greater the indicator, the higher the risk of corruption. Accordingly, sample quartiles are computed and used as thresholds for the risk categories, as follows. Let $X_{ij}$ be the value of red flag $j$ for contracting authority $i$, with $j=\{1,\ldots,15\}{\setminus}\{7,8\}$ and $i=1,\ldots,n$. In addition, let $q_j(0.25), q_j(0.50)$ and $q_j(0.75)$ be the sample quartiles of red flag $j$ (if overlapped, they are computed on the distribution of unique values). The discrete red flag $Y_{ij}$ is obtained as follows:
\[
Y_{ij} = 
\begin{cases} 
0, \qquad \mbox{if } X_{ij} \leq q_j(0.25), \\
1, \qquad \mbox{if } q_j(0.25) <  X_{ij} \leq q_j(0.50), \\
2, \qquad \mbox{if } q_j(0.50) < X_{ij} \leq q_j(0.75), \\
3, \qquad \mbox{if } X_{ij} > q_j(0.75).
\end{cases}
\]

As far as indicators \textit{advertisement} and \textit{evaluation} are concerned, since both extremely low and extremely high values signal a risk of corruption (see Appendix~\ref{appendix}), a different categorisation procedure is carried out. Specifically, the distribution of these indicators is divided into eight parts by computing seven quantiles $q_j(\tau)$, with $\tau$ ranging from 0.125 to 0.875 with step 0.125 (1/8). Afterwards, $Y_{ij}$ is built as follows ($j=7,8$):
\[
Y_{ij} = 
\begin{cases} 
0, \qquad \mbox{if } q_j(0.375) < X_{ij} \leq q_j(0.625), \\
1, \qquad \mbox{if } q_j(0.25) < X_{ij} \leq q_j(0.375) \mbox{ or } q_j(0.625) < X_{ij} \leq q_j(0.75), \\
2, \qquad \mbox{if } q_j(0.125) < X_{ij} \leq q_j(0.25) \mbox{ or } q_j(0.75) < X_{ij} \leq q_j(0.875), \\
3, \qquad \mbox{if } X_{ij} \leq q_j(0.125) \mbox{ or } X_{ij} > q_j(0.875).
\end{cases}
\]
In this way, the low risk condition ($Y_{ij}=0$) includes values of the indicator that lie around the median, whereas the high risk condition ($Y_{ij}=3$) includes values that lie on the extreme tails of the distribution.

The resulting set of $Y_{ij}$ becomes the basis for studying the dimensionality structure of corruption risk in public procurement, by means of the multidimensional GRM introduced in Section~\ref{subsec:model}.

\subsection{Dimensional solutions}\label{subsec:irt_solutions}
The number of dimensions characterising the corruption risk in public procurement is obviously unknown. Consequently, several GRMs are estimated on the data at hand and model fitting is evaluated using classic penalised log-likelihood indexes, as well as model fitting relative improvement computed as the percentage difference of each index with respect to that of the previous model (with one dimension less). Specifically, in this work we consider the following indexes: Akaike Information Criterion \citep[AIC;][]{akaike1973information}, Bayesian Information Criterion \citep[BIC;][]{Schwarz1978}, sample size-adjusted BIC \citep[SABIC;][]{sclove1987application} and Hannan-Quinn information criterion \citep[HQC;][]{hannan1979determination}, together with the classic likelihood ratio test.

The procedure for selecting the suitable number of dimensions begins with fitting the unidimensional model ($D=1$) and increasing until $D=7$. No model is fitted with more dimensions, in order to ideally have sufficient space for allocating at least two red flags in each dimension.

\begin{figure}
    \centering
    \includegraphics[width=\textwidth]{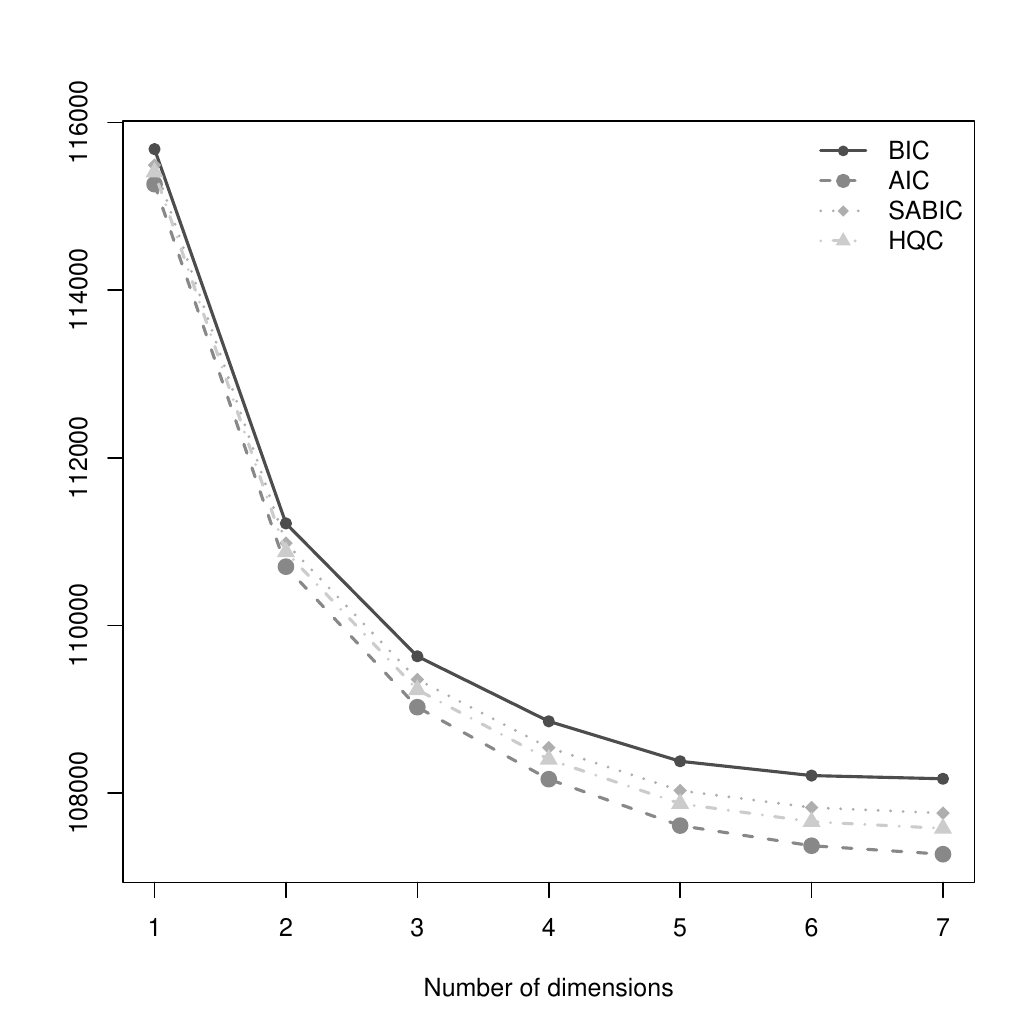}
    \caption{Values of log-likelihood penalised indexes against number of dimensions of the model.}
    \label{fig:model_sel}
\end{figure}

Results (see Table~\ref{tab:model_sel}) highlight $D=7$ as optimal dimensional solution, according to all the indexes and the likelihood ratio test. However, if we consider the relative improvement of the model fitting (index values vs. number of dimensions shown in Figure~\ref{fig:model_sel}), it can be appreciated that the two curves become flat in correspondence of five dimensions, implying that the improvements for models with six or seven dimensions are negligible (lower than~0.2\%), despite higher absolute fitting (so-called elbow method). 

\begin{table}
\caption{Model selection output: number of dimension ($D$), maximimised model log-likelihood ($\hat{l}$), number of model parameters (\# par.), penalised log-likelihood indexes (AIC, SABIC, HQC, BIC), likelihood ratio test statistic ($\chi^2$), degrees of freedom (df) and $p$-value.}
\label{tab:model_sel}
    \begin{tabular}{cccccccrrrr}
    \toprule
    $D$ & $\hat{l}$ & \# par. & AIC & SABIC & HQC & BIC & \multicolumn{1}{c}{$\chi^2$} & df & $p$-value \\
    \midrule
    1 & -57,572.4 & 60 & 115,264.9 & 115,492.7 & 115,408.2 & 115,683.4 & \multicolumn{1}{c}{--}  & \multicolumn{1}{c}{--} & \multicolumn{1}{c}{--} \\
    2 & -55,277.1 & 74 & 110,702.1 & 110,983.2 & 110,878.9 & 111,218.3 & 4,590.7 & 14 & $<0.001$ \\
    3 & -54,426.1 & 87 & 109,026.3 & 109,356.7 & 109,234.1 & 109,633.2 & 1,701.9 & 13 & $<0.001$ \\
    4 & -53,984.6 & 99 & 108,167.1 & 108,543.1 & 108,403.7 & 108,857.7 & 883.1 & 12 & $<0.001$ \\
    5 & -53,696.8 & 110 & 107,613.7 & 108,031.4 & 107,876.4 & 108,381.0 & 575.5 & 11 & $<0.001$ \\
    6 & -53,566.7 & 120 & 107,373.3 & 107,829.1 & 107,660.0 & 108,210.4 & 260.3 & 10 & $<0.001$ \\
    7 & -53,507.2 & 129 & 107,272.4 & 107,762.3 & 107,580.6 & 108,172.2 & 118.9 & 9 & $<0.001$ \\
    \bottomrule
    \end{tabular}
\end{table}

\begin{table}
    \caption{Output of the IRT model with $D=5$ dimensions: oblimin-rotated loadings (cut at $\pm$0.2), communalities ($h^2$), sum of squared (SS) loadings and proportion of explained variance by each dimension.}
    \label{tab:sel5}
    \centering
        \begin{tabular}{lcccccc}
        \toprule
        Red flag & D1 & D2 & D3 & D4 & D5 & $h^2$ \\
        \midrule
        \textit{non\_open\_count} & 0.924 &  &  &  &  & 0.837 \\
        \textit{non\_open\_val} & 0.983 &  &  &  &  & 0.990 \\
        \textit{single\_bid\_count} &  & 0.925 &  &  &  & 0.875 \\
        \textit{single\_bid\_val} &  & 0.970 &  &  &  & 0.936 \\
        \textit{MEAT\_count} &  &  & 0.944 &  &  & 0.860 \\
        \textit{MEAT\_val} &  &  & 0.978 &  &  & 0.993 \\
        \textit{advertisement} &  &  &  &  &  & 0.031 \\
        \textit{evaluation} &  &  &  &  &  & 0.015 \\
        \textit{excluded\_bids} &  & -0.232 &  & 0.794 &  & 0.695 \\
        \textit{all\_bids\_excluded\_but\_one} &  &  &  & 0.962 &  & 0.932 \\
        \textit{excluded\_bids\_but\_one} &  &  &  & 0.830 &  & 0.696 \\
        \textit{modifications} &  &  &  &  & 0.928 & 0.891 \\
        \textit{amount\_deviation} &  &  &  &  & 0.699 & 0.472 \\
        \textit{time\_deviation} &  &  &  &  & 0.356 & 0.163 \\
        \textit{winners\_homog} &  &  &  &  &  & 0.009 \\
        \midrule
        SS loadings & 1.863 & 1.498 & 1.905 & 2.261 & 1.848 & \\
        Explained variance (prop) & 0.124 & 0.100 & 0.127 & 0.151 & 0.123 &  \\
        \bottomrule
        \end{tabular}
\end{table}

By concurrently accounting for both a statistical criterion based on the above indexes and a criterion that is attentive to model parsimony~\citep{latentmarkov, pohle2017selecting, montanari2018multilevel}, the solution with $D=5$ is selected as the best one. Table~\ref{tab:sel5} reports the five-dimensional solution in terms of rotated loadings, communalities, sum of squared loadings, and proportion of explained variance. These results align with the exploratory analysis, based on correlation among red flags (Section~\ref{subsec:exploratory_analysis}). In fact, the first three pairs of red flags -- related to the proportion of non open procedures, to the procedures receiving a single bid, and to the procedures awarded with the MEAT criterion -- contribute to defining three distinct dimensions (D1, D2, and D3) due to high loadings and, consequently, the variability of these indicators is almost entirely explained (high communalities). Two further groups of red flags emerge, each defining a different dimension. The first group considers bid exclusion (\textit{excluded\_bids}, \textit{all\_bids excluded\_but\_one} and \textit{excluded\_bids\_but\_one}), which flows into dimension D4. The second group includes indicators based on variants and the deviation in contract economic values and duration (\textit{modifications}, \textit{amount\_deviation}, and \textit{time\_deviation}), contributing to dimension D5.

Finally, indicators concerning the average time for submitting bids (\textit{advertisement}) and for selecting winning bids (\textit{evaluation}), together with the indicator on the homogeneity of the distribution of winners (\textit{winners\_homog}), do not consistently load into none of the five ascertained dimensions and show very low communalities, thus highlighting that these dimensions can explain only a very low share of their variability. Again, this is an expected result, since these three red flags have very low correlation coefficients with the other indicators (Section~\ref{subsec:exploratory_analysis}). 
We have also fitted the model with six dimension to confirm if excluded indicators are re-admitted to a six-dimensional solution, but none contribute to an additional risk dimension measurement.

\begin{table}
\caption{Output of the IRT model with $D=4$ dimensions: oblimin-rotated loadings (cut at $\pm$0.2), communalities ($h^2$), sum of squared (SS) loadings and proportion of explained variance by each dimension.}
\label{tab:sol4}
\centering
    \begin{tabular}{lccccc}
    \toprule
    Red flag & D1 & D2 & D3 & D4 & $h^2$ \\
    \midrule
    \textit{non\_open\_count} & 0.918 &  &  &  & 0.837 \\
    \textit{non\_open\_val} & 0.981 &  &  &  & 0.991 \\
    \textit{single\_bid\_count} &  & 0.908 &  &  & 0.839 \\
    \textit{single\_bid\_val} &  & 0.991 &  &  & 0.971 \\
    \textit{MEAT\_count} &  &  & 0.940 &  & 0.862 \\
    \textit{MEAT\_val} &  &  & 0.973 &  & 0.993 \\
    \textit{advertisement} &  &  &  &  & 0.026 \\
    \textit{evaluation} &  &  &  &  & 0.004 \\
    \textit{excluded\_bids} &  &  &  & 0.766 & 0.649 \\
    \textit{all\_bids\_excluded\_but\_one} &  &  &  & 0.978 & 0.957 \\
    \textit{excluded\_bids\_but\_one} &  &  &  & 0.832 & 0.707 \\
    \textit{modifications} &  &  & -0.252 & 0.261 & 0.135 \\
    \textit{amount\_deviation} &  &  &  & 0.434 & 0.213 \\
    \textit{time\_deviation} &  &  &  & 0.284 & 0.090 \\
    \textit{winners\_homog} &  &  &  &  & 0.007 \\
    \midrule
    SS loadings & 1.959 & 2.575 & 1.880 & 1.880 &  \\
    Explained variance (prop) & 0.131 & 0.172 & 0.125 & 0.125 & \\
    \bottomrule
    \end{tabular}
\end{table}

Besides, the solution with four dimensions (Table~\ref{tab:sol4}) is also inspected to assess how the final five dimensions get clustered. We find again the three blocks of two indicators (D1, D2, and D3), while the remaining six red flags are clustered in D4. The indicators about the exclusion of bids have high loadings, while the proportion of procedures with variants has similar magnitudes on D3 and D4 (even if of opposite sign). Likewise for the five-dimension solution, the same three indicators (\textit{advertisement}, \textit{evaluation} and \textit{winners\_homog}) remain outside the four-dimensional solution.

Given all the previous considerations, we retain the five-dimensional solution as the most suitable for characterising corruption risk in public procurement. 
The correlation matrix among the five dimensions (not shown) displays poor correlations among them. The only exception involves dimensions D1 (related to non-open procedures) and D3 (MEAT criterion recurrence), which are are negatively correlated ($-$0.458), as, by the way, already shown in Figure~\ref{fig:corrplot}.

Finally, for each red flag overall discrimination and difficulties (IRT parameters) are obtained using formulas in~(\ref{eq:overall_discr}) and~(\ref{eq:overall_diff}) on the estimated parameters of the five-dimensional model. Looking in particular at the discriminating power of each red flag and considering the underlying corruption risk dimensional structure (Table~\ref{tab:irt_params}), we can appreciate how red flags based on contract economic value have the highest overall discrimination parameter, both within and across dimensions. In this perspective, other important red flags include the one that considers procedures with all bids excluded but one and the indicator that takes into account modified contracts.

\begin{table}
\caption{Estimated IRT parameters for each red flag $j$: overall discrimination $\hat{\alpha}_j$ and difficulties $\hat{\beta}_{jy}$. }
\label{tab:irt_params}
\centering
    \begin{tabular}{clr@{\qquad}rrr}
    \toprule
     & Red flag & $\hat{\alpha}_j$ & $\hat{\beta}_{j1}$ & $\hat{\beta}_{j2}$ & $\hat{\beta}_{j3}$ \\
    \midrule
    \multirow{2}{*}{D1} & \textit{non\_open\_val} & 16.56 & -0.89 & -0.51 & -0.14 \\
    & \textit{non\_open\_count} & 3.86 & -1.04 & -0.40 & 0.05 \\
    & \\
    \multirow{2}{*}{D2} & \textit{single\_bid\_val} & 6.53 & 0.26 & 0.63 & 1.05 \\
    & \textit{single\_bid\_count} & 4.51 & 0.03 & 0.40 & 0.77 \\
    & \\
    \multirow{2}{*}{D3} & \textit{MEAT\_val} & 20.91 & -0.08 & 0.30 & 0.67 \\
    & \textit{MEAT\_count} & 4.24 & -0.28 & 0.26 & 0.82 \\
    & \\
    \multirow{3}{*}{D4} & \textit{all\_bids\_excluded\_but\_one} & 6.31 & 1.01 & 1.16 & 1.43 \\
    & \textit{excluded\_bids\_but\_one} & 2.58 & 1.84 & 2.01 & 2.54 \\
    & \textit{excluded\_bids} & 2.57 & 0.02 & 0.54 & 1.17 \\ 
    & \\ 
    \multirow{3}{*}{D5} & \textit{modifications} & 4.86 & -0.08 & 0.15 & 0.54 \\
    & \textit{amount\_deviation} & 1.61 & 0.11 & 0.78 & 1.57 \\
    & \textit{time\_deviation} & 0.75 & -0.64 & 0.67 & 2.15 \\
    & \\
    \multirow{3}{*}{--} & \textit{advertisement} & 0.30 & -4.37 & -0.63 & 2.90 \\
    & \textit{evaluation} & 0.21 & -5.42 & -0.19 & 4.93 \\
    & \textit{winners\_homog} & 0.16 & 5.37 & 8.06 & 12.36 \\
    \bottomrule
    \end{tabular}
\end{table}

\section{Discussion of main results}\label{sec:concl}
In this paper we assessed the validity of a set of red flag indicators through a procedure based on the framework of multidimensional IRT models, which allows us to account concurrently for the latent and complex facet of corruption. 
A pool of fifteen red flag indicators of corruption risk in public procurement was chosen and computed on data included in the Italian National Database of Public Contracts (BDNCP). Results showed that the risk of corruption is not a one-dimensional phenomenon, because the goodness of fit of the multidimensional models to the data was always better than that observed for the one-dimensional solution. This result is fully in line with our expectations as corruption risk is a complex phenomenon that occurs through behaviours, facts and circumstances that differ in nature, type and entity. Operationally, this result implies that the single measures of corruption risk (i.e., red flags) are not expected to be all related with each other, but rather in sub-groups, each measuring a different and specific risk category of the same underlying risk of corruption. Indeed, among the several multidimensional solutions we identified, the five- and four-dimensional solutions are to be preferred, on the basis of statistical, interpretability and model parsimony criteria. In the five-dimensional solution, the five categories of corruption risks refer in particular to: 
\begin{enumerate}
    \item risks related to the type of procurement procedure, or to the weight -- in relative terms and in terms of economic value -- of non-open procedures (for example, direct assignments or negotiated procedures) on the total number of procedures awarded; 
    \item risks related to the circumstance that a procurement procedure receives only one offer, or to the weight -- in relative terms and in terms of economic value -- of the procurement procedures that receive only one offer; 
    \item risks related to the contract award criterion, or to the weight -- in relative terms and in economic value -- of the procedures awarded with discretionary criteria, such as that of the most economically advantageous tender; 
    \item risks related to the circumstance that, during the evaluation of the offers received, a contracting authority excludes all but one offers, or to the weight in relative terms of the procurement procedures for which all the offers have been excluded except for one; 
    \item risks related to the circumstance that a contract undergoes variations during its implementation and to the extent of the variation between $i.$ allotment amount and cleared, and $ii.$ contract expected and actual duration.
\end{enumerate}

The risk of corruption measured by the first and third categories are greatly related to an excess of discretionary power exerted by contracting authorities in the public procurement decision process.
Addressing excess of discretion is key to advance socioeconomic sustainability, which requires transparency and accountability in decision-making. An excess of discretion undermines these principles by allowing contracting authorities to make decisions without clear guidelines and oversight. Even though access to (and availability of) relevant information related to the public procurement decision-making process may still be formally guaranteed by contracting authorities, excess of discretionary power can weaken disclosure and accountability mechanisms (i.e., public consultations, complaint mechanisms, audit processes, etc.) and lead to favouritism, conflict of interest and the misuse of resources. Unwarranted discretion can also erode the rule of law and undermine fairness in socio-economic systems. Indeed, when decisions are based on subjective judgement rather than objective criteria, fairness -- a fundamental pillar of socio-economic sustainability, ensuring equal opportunities, protection of rights, and access to resources for all individuals and groups -– may be compromised. The two above groups of red flags are therefore of particular interest of many targets under SDG~16 and beyond, such as (among others): Target 16.6 devoted to developing effective, accountable and transparent institutions; Target 16.3 devoted to promoting the rule of law and equal access to justice; Target 10.3 devoted to ensuring equal opportunity and reduce inequalities of outcome, by eliminating discriminatory laws, policies and practices.

The second and fourth categories identify in the lack of competitiveness the major mechanism behind corruption risks. Both dimensions point to the absence of competitive bids as a sign that collusive practices may have taken place among contracting authorities and companies, i.e., in the form of illegal agreements to ensure that a specific bidder wins the contract. Competition plays a crucial role in fostering economic growth, innovation, and equitable outcomes. Conversely, lack of competition can concentrate wealth and power in the hands of a few dominant players, exacerbating socio-economic inequalities and hindering sustainable development through higher prices, reduced consumer welfare, lower product diversity and limited innovation. Hence, these two aggregations of red flags relate especially to SDG targets on issues conveying sustainable economic growth and innovation, such as: Target 8.2 on achieving higher levels of economic productivity through diversification, technological upgrading and innovation; Target 8.8 which calls for promoting a stable and secure working environment, and implies a fair competition environment where businesses can thrive based on merit and innovation; Target 9.5 devoted to enhancing scientific research and innovation. Besides, the two aggregations may be of interest transversely of those SDGs and targets devoted to reducing inequalities, such as Goal 10 and Goal 5 devoted specifically to achieve gender equality.

Finally, the fifth group of red flags primarily reflects inefficiencies in the public procurement process. Efficiency is a fundamental concept throughout the 2030 Agenda, as it is closely linked to achieving sustainability and ensuring the responsible allocation and use of resources. This last group of red flags is therefore of interest of several SDGs and related targets, such as, among others: Target 9.4, which focuses on upgrading infrastructure and retrofitting industries to make them sustainable, with a particular emphasis on resource efficiency and adoption of clean and environmentally sound technologies; Target 12.2, which calls for achieving sustainable management and efficient use of natural resources.

Besides, the analysis of the correlation between sub-dimensions of corruption risk shows negligible (and close to zero) correlations between dimensions. The identified dimensions of the risk of corruption are therefore generally non-superimposable, and different from each other. Based on our findings, this implies in particular that: 

\begin{itemize}
    \item the risk associated with the adoption of non-open procedures is neither related with the risk associated with the circumstance that a procurement procedure receives only one offer, nor with the risk associated with the circumstance that all offers, except one, are excluded; 
    \item  the risk associated with the circumstance that a tender procedure receives only one offer is not related with the risk associated with the circumstance that a tender procedure is awarded with discretionary criteria, nor with the circumstance that, during the evaluation phase of the offers received, a contracting station excludes all but one offer;
    \item the risk linked to the circumstance that a tender procedure is awarded with discretionary criteria is not associated with the risk linked to the exclusion of all offers except one.
\end{itemize}

The only important exception was given by the negative and by no means negligible correlation between the risk linked to the adoption of non-open procedures and the risk linked to the circumstance that a tender procedure is awarded with discretionary criteria. This implies that non-open procedures are generally also those which are awarded through criteria other than the discretionary ones, such as the criterion of the lowest price. Finally, from the five-dimensional solution it is possible to observe that the risk linked to the circumstance that a contract undergoes variations during implementation is negatively associated (albeit with a weak entity) both to the risk that the tender procedure receives a single offer, and to the risk linked to the circumstance that a tender procedure is awarded with discretionary criteria.

The last analysis concerned the estimation of the discriminating power of the red flag indicators within each sub-dimension of corruption risk. Since discrimination estimates can be interpreted as weights assigned to the single red flags, they help in identifying which red flag indicators contribute the most to differentiating units of analysis within each sub-dimension. In both the four- and five-dimensional solutions, the most important red flags were those accounting for the economic value of tendering procedures, and specifically: economic value of procedures awarded with discretionary criteria; economic value of unopened procedures; economic value of tender procedures that receive a single offer; bidding procedures in which all bids, but one, are excluded. 

The present work can be extended in several ways. First, in order to check the robustness of the obtained results across time and space, the proposed validating procedure can be tested with data referred to other years and to other national contexts. Second, a multilevel component accounting for the nested structure of our data (i.e., contracts nested within contracting authorities) can be added to the proposed statistical model to get insights on the contribution of data gathered at different aggregation levels to the identified dimensionality solutions. 

\appendix

\section{Description of the red flags}\label{appendix}

In this section, we report a brief description of each selected red flag (see Tables~\ref{tab:sel_redflags1}, \ref{tab:sel_redflags2}, \ref{tab:sel_redflags3}), along with the related relevant literature.

\subsubsection*{Proportion of non-open procedures (number and value)}
Lack of competition is considered a high-risk condition for corruption in procurement procedures \citep{abdou2021covid, klavsnja2015corruption, wachs2021corruption, fazekas2020uncovering, fazekas2016objective, fazekas2016comprehensive, fazekas2022extra}. Non open and non competitive procedures, such as direct awards and negotiated procedures, are therefore considered risky procedures. Accordingly, this red flag is computed accounting for the incidence of non open procedures over the total number of tenders managed by a contracting authority. At the contract-level, the red flag is computed as a dummy variable assuming value 1 when the procedure is non open, and 0 otherwise. When computed at the contracting body level, the red flag becomes a proportion of non open procedures out of all the procedures handled by a contracting authority. The first of the two red flags named \textit{non\_open\_count} indicates, for each contracting authority, the proportion of non open procedures out of all the procedures. The \textit{non\_open\_val} red flag is computed by dividing the economic value of non open tenders out of the total economic value of all tenders. The latest red flag can be considered a weighted version of the former, where the weight is expressed by the economic value of the tender. 
 
The literature \citep{auriol2016public, anac2017, decarolis2019corruption, abdou2021covid, fazekas2022extra, olaf2017} underlines that these red flags may not necessarily signal corruption since the use of non competitive procedures are allowed in most current regulations. However, a high percentage of tenders awarded according to non competitive mechanisms could signal the existence of risky circumstances that needs to be specifically monitored.

\subsubsection*{Proportion of procedures with a single bid (number and value)} One of the most striking cases of lack of competition in the public procurement process occurs when a procurement procedure receives only one offer~\citep{fazekas2016objective, fazekas2020uncovering}. The red flag at issue therefore detects the incidence of procedures receiving a single bid over all tenders managed by a contracting authority. Similarly to the previous indicator, the present red flag is computed at the contract level as a dummy variable assuming value 1 when the procurement procedure receives a single offer, and 0 otherwise. When computed at the contracting body level, it becomes a proportion of tenders receiving a single offer. Again, two versions are available: the unweighted indicator and its weighted version by contract economic value.

\subsubsection*{Proportion of procedures awarded with the most economically advantageous tender criterion (count and value)} 
Contracting authorities can award public contracts according to two main assessment criteria, the lowest price and the most economically advantageous tender (MEAT). Through the MEAT criterion, contracting bodies compares bids to the best value for money. The MEAT is considered a proxy for corruption risk \citep{anac2017, decarolis2019corruption, fazekas2020uncovering, olaf2017}, since discretionary criteria (e.g., assessment of technical merit, aesthetic and functional characteristics, originality, etc.) may come into play in the assessment of the contract quality and may limit accountability control mechanisms. Like the two previous red flags, this indicator is binary at the contract level (presence/absence of MEAT), while it is a proportion of MEAT contracts when considering the contracting authority aggregation. Again, unweighted/weighted by contract economic value versions are computed.

\subsubsection*{Average time for submitting bids and for selecting the winning bid}
These red flags consider two time spans. The first measures the time interval between the date of publication of the call for tenders and the deadline for the submission of bids. In fact, the literature \citep{olaf2013, fazekas2020uncovering, fazekas2016comprehensive, fazekas2016objective, olaf2017} agrees that extremely short advertisement periods may inhibit non-connected bidders to develop adequate bids in time. In parallel, lengthy periods may hide legal issues.
 
The second indicator measures the time span between the deadline for the bid submission  and the award notice date. It indicates the time interval to assess the bids and award the contract and is considered a risk measure \citep{olaf2013, fazekas2020uncovering, fazekas2016comprehensive, fazekas2016objective, olaf2017} as too short time intervals to award contracts and snap decisions may imply premeditated appraisal. Again, also a long evaluation period is seen as a signal of corruption, as it may mirror legal challenges in the selection of winners. 

When computed at the contracting body level, these red flags consist in the average time span (in days) between the 
relevant dates.

\subsubsection*{Proportion of excluded bids}
It is broadly acknowledged that a further risky condition occurs whenever the number of bids is limited as a result of an exclusion made by the contracting authority during the evaluation stage \citep{olaf2013, olaf2017, scomparin2017corruzione, ferraris2016warning, anac2017, ferwerda2017corruption, fazekas2016objective}. The underlying assumption is that the greater the proportion of excluded bids, the higher the risk of corruption. 
Indeed, the exclusion of a large proportion of bids can signal a selection strategy by the contracting body aimed at favouring pre-selected companies.
 
At the single contract level, this red flag considers the proportion of excluded bids out of the total number of bids received for that procedure. When computed at the contracting authority level, it is obtained as the average proportion of excluded bids over all procedures managed by that authority. 

\subsubsection*{Proportion of procedures with all bids excluded but one}
The indicator detects the fraction of procedures for which all bids but one have been excluded during the evaluation phase, out of the total number of contracts. Again, the exclusion of all bids but one, just like the cases where only one bid is submitted, is considered a risky condition because it implies absence of competition in the public procurement procedure~\citep{fazekas2016objective}. Similarly to the previous red flags, for each contract a dummy variable is created which assumes value 1 when all bids but one are excluded and 0 otherwise. The red flag at issue becomes a proportion of tenders with this feature when computed at the level of contracting authorities.

\subsubsection*{Proportion of excluded bids in procedures with all bids excluded but one}
This red flag consists in a modification of indicator related to the proportion of excluded bids and, at the contracting body level, it measures the average proportion of excluded bids over the procedures for which all bids but one have been excluded. The indicator allows us to distinguish two opposite cases with $i.$ high number of bids, where the exclusion aimed at leaving only one bid concerns a large number of bids (e.g., 15 submitted bids with the subsequent exclusion of 14 of them), and $ii.$ low number of bids, in which the exclusion concerns a limited number of bids (e.g., 4 submitted bids with the exclusion of 3 of them). It is supposed that the risk of corruption is higher in the first of the two circumstances outlined above.

\subsubsection*{Proportion of procedures with variants}
This indicator measures the fraction of awarded contracts modified through variants during their execution as in certain circumstances, this may signal a pathology to monitor \citep{anac2017, olaf2013, olaf2017}. At the single contract level, the red flag is computed as a binary variable reporting whether at least one variant has occurred. At the contracting body level, the red flag is computed as a proportion of modified contracts out of all concluded contracts. 

\subsubsection*{Deviations in the contract economic value and duration}
These red flags consider two deviations between actual and expected quantities. Specifically, the first (\textit{amount\_deviation}) considers whether there has been a deviation of the contract actual execution economic value from its initial (and expected) awarded value. It measures, at the single contract level, the relative distance between actual sums paid by the contracting body and the awarded value. When computed at the level of single administrations, it is obtained by averaging the deviations over contracts managed by each contracting body. The indicator can be useful to assess possible \virg{moral hazard} behaviours during contract execution, as it happens when companies apply very high discounts in the awarding phase and then recover them during the execution phase. Even though unexpected circumstances could lead to legitimate increases of execution costs, the literature \citep{anac2017, fazekas2022extra, olaf2013, olaf2017} agrees that variants may hide connivance between companies and contracting authorities to artificially increase contract costs. 
 
At the single contract level, the indicator \textit{time\_deviation} focuses on the deviation (again, as a relative distance) of actual execution times with respect to those expected by the contract. When computed for the single administration, it is obtained by averaging the deviations over the contracts managed by the contracting body. Similarly to the previous, this indicator is intended to assess opportunistic company behaviours indulged by contracting authorities, even though, again, time deviations may be justified by legitimate suspensions \citep{anac2017, decarolis2019corruption, fazekas2022extra, fazekas2020uncovering}.

\subsubsection*{Homogeneity of winners' distribution}
The indicator is a modified version of a well-known indicator named \virg{Winner's share of issuer contracts}. In its original version \citep{fazekas2020uncovering, fazekas2022extra, abdou2021covid}, this red flag measures the proportion of contract value awarded by a contracting station to a winning company relative to the total value of all contracts awarded by the same contracting station. Our proposal considers the number of procedures rather than the economic awarded value. The proposed indicator therefore assesses the recurrence (or frequency) with which a contracting authority awards its contracts to the same company(ies). The literature -- see, as an example, the policy reports from international bodies such as the World Bank Group\footnote{\url{https://www.worldbank.org/en/about/unit/integrity-vice-presidency/annual-reports}} -- often recalls the risk condition associated with the consecutive awarding of contracts to the same companies, which can find justification only if it entails an economic advantage. 
 
This red flag can be computed by measuring the degree of homogeneity of the frequency distribution of winner firms for a particular contracting body. For this purpose, let us consider contracting authority $i$, whose contracts are awarded to, say, $K_i$ companies and let $f_ {ik}$ be the proportion of contracts awarded to company $k$, with $k=1,\ldots,K_i$.

The least at risk situation occurs when $f_{ik}=1/K_i\: \forall k$, that is, the contracting authority has the same award rate over their winner companies, which corresponds to a situation of maximum heterogeneity, or minimum homogeneity, in the frequency distribution. As heterogeneity decreases (then homogeneity increases), the concentration of contracts awarded to the same company(ies) increases. Hence, a homogeneity index for contracting authority~$i$ -- which represents the red flag \textit{winners\_homog} -- can be obtained through the classic heterogeneity index of a frequency distribution \citep[Chapter 5]{cicchitelli2012statistics}:
\[
H_i = 1 - \frac{K_i}{K_i-1} \Bigl( 1 - \sum_{k=1}^{K_i} f_{ik}^2 \Bigr ).
\]
This index ranges between 0 (minimum homogeneity, maximum heterogeneity) and 1 (maximum homogeneity, minimum heterogeneity).


\bibliographystyle{apalike}
\bibliography{sn-bibliography}

\end{document}